\documentclass[manuscript]{aastex}
\usepackage{natbib}                                        
\usepackage{graphicx}                                     
\usepackage{amsmath,amssymb}
\usepackage{color}                              

\newcommand{\pd}[2]{\frac{\partial #1}{\partial #2}}       
\newcommand{\pdd}[2]{\frac{\partial^2 #1}{\partial #2^2}}  
\newcommand{\td}[2]{\frac{d #1}{d #2}}                    
\newcommand{\tb}[1]{\mathbf{#1}}                          

\shorttitle{Modeling the Sunspot Number Distribution} 
\shortauthors{Noble and Wheatland}
\slugcomment{Accepted by the Astrophysical Journal 25 Feb 2011}
\begin{document}

% title etc
\title{Modeling the Sunspot Number Distribution with a Fokker-Planck Equation}
\author{P.~L. Noble and M.~S. Wheatland}
\affil{Sydney Institute for Astronomy, School of Physics, The University of Sydney, NSW 
2006, Australia}
\email{p.noble@physics.usyd.edu.au}

% abstract 
\begin{abstract}
Sunspot numbers exhibit large short--timescale (daily--monthly) variation in addition to longer timescale variation due to solar cycles.  A formal statistical framework is presented for estimating and forecasting randomness in sunspot numbers on 
top of deterministic (including chaotic) models for solar cycles. The Fokker--Planck 
approach formulated assumes a specified long--term or secular variation in sunspot number over an underlying solar cycle via a driver function.  The model 
then describes the observed randomness in sunspot number on top of this driver function.  We 
consider a simple harmonic choice for the driver function, but the approach is general and can easily 
be extended to include other drivers which account for underlying physical processes 
and/or empirical features of the sunspot numbers.  The framework is consistent 
during both solar maximum and minimum, and requires no parameter restrictions to ensure non-negative sunspot numbers.  Model parameters are estimated using statistically optimal 
techniques.  The model agrees both qualitatively and quantitatively with monthly sunspot 
data even with the simplistic representation of the periodic solar cycle.  This framework should be particularly useful for solar cycle forecasters and is complementary to existing modeling techniques.  An analytic approximation for the Fokker--Planck equation is presented, which is analogous to the Euler approximation, which which allows for efficient maximum likelihood estimation of large data sets and/or when using difficult to evaluate driver functions.

\end{abstract}

\section{Introduction}
Sunspots are regions on the solar surface where
strong magnetic fields pierce the photosphere.  Sunspots form when
magnetic flux tubes rise out of the solar interior and cross the
photosphere.   The magnetic field around these flux tubes is
sufficiently strong to disrupt the usual process of convective heat
transport to the surface, so sunspots appear as small dark spots on the
photosphere.  Areas around sunspots where surface magnetic fields are
particularly strong are known as active regions and host explosive
magnetic events including solar flares and coronal mass ejections
\citep{EmslieEtAl1998}.

The daily sunspot number is described using the Wolf number 
\begin{equation} s = k \left( 10g + N \right),
\end{equation}
where $g$ is the number of sunspot groups and $N$ is the number of
individual spots \citep{BruzakEtAl1977}.  The sunspot number is a
weighted average of individual spots and groups, and the correction factor $k$
depends on a number of factors, including the location of the observatory,
instrument parameters, and counting method \citep{Petrovay2010}.  The data used in
this paper is a weighted average of measurements from a network of
observatories (the `International sunspot number'), produced by the
Solar Influences Data Analysis Center (SIDC), Royal Observatory of
Belgium\footnote{Sunspot data is provided by the US National
Geophysical Data Center (NGDC) at http://www.ngdc.noaa.gov/stp/spaceweather.html.}.  The sunspot number is a constructed measure of
solar activity and not a physical quantity, and is represented by a
non--negative number with an important lower limit of zero corresponding
to no visible active regions.

The mean sunspot number varies with a semi--regular 11--year 
cycle \citep{Parker1955}, but there are large daily, weekly, and yearly fluctuations on 
top of this regularity, as well as large variations in the maximum sunspot number in 
a cycle.  The magnetic fields responsible for the formation of active regions are generated by a dynamo process in the solar interior \citep{Tobias2002}.  In the solar convection zone, the flow of plasma and magnetic fields is turbulent due to the high value of the hydrodynamic Reynolds number \citep{Ossendrijver2003}.  As a result it is difficult to accurately model the strength, location, and timing of magnetic fields appearing at the solar surface \citep{Choudhuri2008}.  A large body of literature suggests that the underlying solar cycles driven by the dynamo also involve chaotic dynamics \citep{LetellierEtAl2006, HanslmeierEtAl2010, AguirreEtAl2008}.

Comprehensive reviews of techniques for modelling/forecasting solar cycles have been presented by \cite{Kane2007}, \cite{Pesnell2008}, and \cite{Petrovay2010}.  Existing methods generally use averaged/smoothed sunspot data, which means they describe the underlying solar cycle and not short--term fluctuations in the sunspot number \citep{Petrovay2010}.  This is an important difference.  Substantial day--to--day fluctuations occur due to both the rapid appearance/formation of large active regions and fast development of magnetic structures within active regions.  These drive extreme space weather events which affect the Earth \citep{2008sswe.rept.....C}.  On average, the sunspot number will jump by more than 50 in a single day more than 20 times per solar cycle. The largest single--day jump for the interval 1850--2010 was $\Delta s = 112$, which occurred in April 1947.  The characteristics of long--term variations in solar cycles have been extensively studied, but the distribution of these large short--term fluctuations has not.  In this paper we introduce a formal statistical framework for modeling day--to--day fluctuations in sunspot number.  Our approach is similar to a recent paper by \cite{AllenEtAl2010} in that it treats the sunspot numbers as a diffusion process, but it has a number of specific advantages over this earlier method.  First, our model provides a framework for combining deterministic (including chaotic) models for secular variation in solar cycles with statistical analysis of the sunspot number time series.  Hence the framework could in principle incorporate chaotic--oscillator type models to account for pseudo--periodic solar cycles underneath short--timescale stochastic fluctuations in sunspot number.  Second, the model allows model parameters to be estimated from the data using maximum likelihood (ML), which provides optimal estimates in the sense of efficiency and consistency in large samples \citep{DacunhaCastelleEtAl1986}.  Third, the formulation enforces the non--negativity of sunspot number, so that the model is valid during both solar maximum and minimum.  This avoids ad hoc treatment of the boundary condition at zero sunspot number, a problem with the \cite{AllenEtAl2010} method.   

The layout of this paper is as follows.  In section \ref{Section:FPE} we derive a Fokker--Planck equation for the sunspot number distribution, and illustrate the general 
properties of the equation using a toy model with a simple harmonic choice for the 
periodic driver function.  Section 3.1 summarizes the details of maximum likelihood 
estimation of diffusion processes.  In section \ref{Section32} we apply the ML technique 
to monthly sunspot data for the interval 1975 to 2009, again using a simple harmonic choice of driver function.  The results agree both qualitatively and quantitatively with the empirical sunspot distribution.  Section \ref{Section4} presents an analytic approximation of the Fokker--Planck equation (\ref{eq:GeneralFPE}), which allows for efficient maximum likelihood estimation of large data sets and/or when using driver functions which are difficult to evaluate. 

% Section -> Sunspot Data
\section{A Fokker--Planck equation for the sunspot number}
\label{Section:FPE}
In this section we introduce a continuous--time stochastic model for sunspot number using 
a Fokker--Planck equation.  Sunspots form and disappear on the visible solar 
surface continuously, so it is intuitive to represent the sunspot number at time $t$ 
with a continuous variable $s(t)$.  Due to complicated physical processes associated 
with sunspot formation and evolution, the sunspot number is uncertain and $s(t)$ is 
stochastic.  As such, we are interested in the time evolution of the probability 
distribution function (pdf) of the sunspot number given an initial sunspot number $s
(t_0)=s_0$ at time $t_0$.  We denote this conditional pdf
\begin{equation}
f(s,t) = f(s,t|s_0),
\end{equation}
where $f(s,t)ds$ is the probability that $s(t)$ lies in the range $\left( s,s + ds 
\right)$ at time $t$ given that it was initially at $s_0$.

Long term or secular variation in the sunspot numbers due to the solar cycle is represented by a driver function $\theta(t)$.  This driver function is chosen to reflect underlying physical processes and/or empirical features of the solar cycles (e.g. a semi--periodic dynamo, the Gnevyshev Gap \citep{Gnevyshev1967}, the Waldmeier effect \citep{Waldmeier1935}, asymmetric/chaotic cycles etc).  For example, the function $\theta(t)$ might be the solution to a system of nonlinear differential equations, in which case the model could describe chaotic solar cycles.  The model presented here does not attempt to account for the solar cycle, which must be contained in the choice of a periodic function for $\theta(t)$.  The model describes the randomness on top of this 
underlying secular variation.

\subsection{Statistics of sunspot data and motivation}
\label{Section21}
\begin{figure}[ht]   
\epsscale{1.0}
\plotone{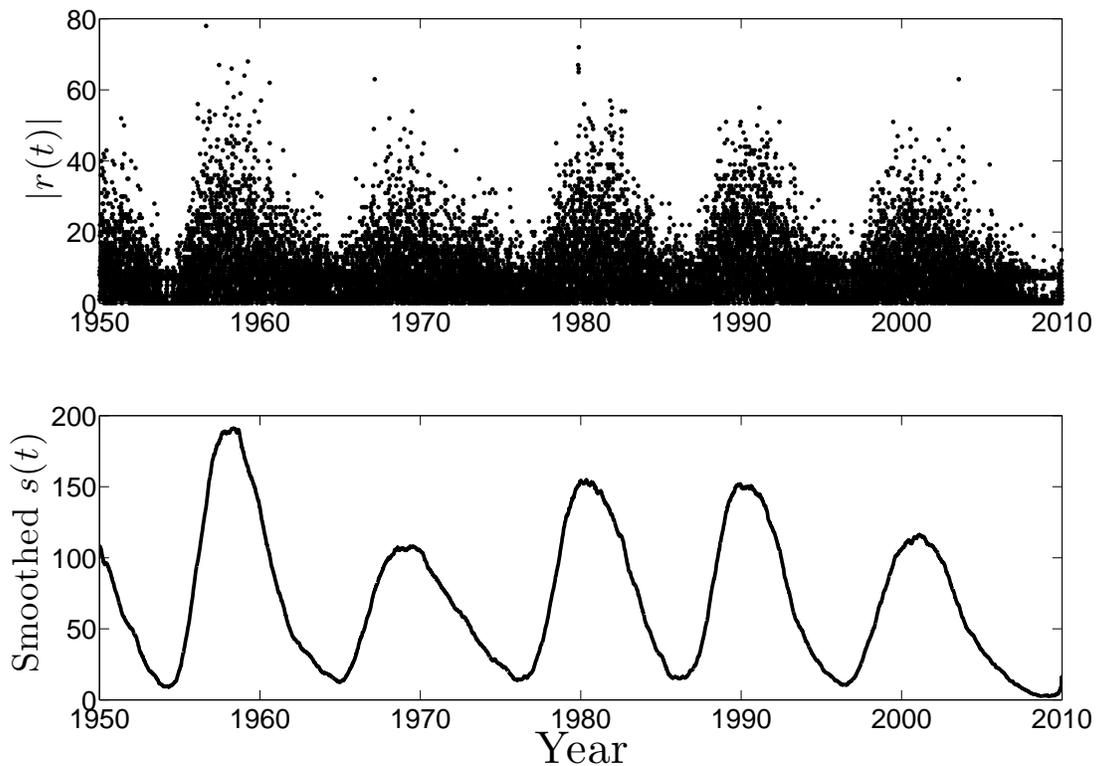}
\caption{Figure showing how the the variance in sunspot numbers increases with sunspot number.  The upper panel shows the absolute deviations $|r(t)| = |s(t) - s(t-\Delta t)|$ between consecutive daily sunspot numbers over the last 60 years.  The lower panel is a smoothed sunspot number time series showing the underlying solar cycle.  A minimal model for short--term fluctuations in $s(t)$ must describe the non--zero variance at zero sunspot number, and the observed increase in variance with sunspot number.}
\label{fig:SSNVariance}
\end{figure}

To some insight into the short--term fluctuations in sunspot number on top of the solar cycle variation, we consider the empirical distribution of daily sunspot numbers.  The size of deviations between consecutive observations of the sunspot number data $|r(t)| = |s(t)-s(t-\Delta t)|$, where $\Delta t$ is the daily time step in the observations, is a proxy for the standard deviation (so that $r(t)^2$ corresponds to the variance) in sunspot number at different times during the solar cycle.  Figure \ref{fig:SSNVariance} shows that this quantity increases with the solar cycle.  The upper panel plots $|r(t)|$ over the last sixty years.  The lower panel is a smoothed daily sunspot number time series showing the underlying solar cycle over the same period.  A minimal description of this data requires an account of the underlying solar cycle (here provided by the driver function $\theta(t)$), as well as a statistical model accounting for the observed non--zero variance at zero sunspot number, and the observed increase in variance with the amplitude of the underlying solar cycle.   

\begin{figure}[ht]   % colour
\epsscale{1.0}
\plotone{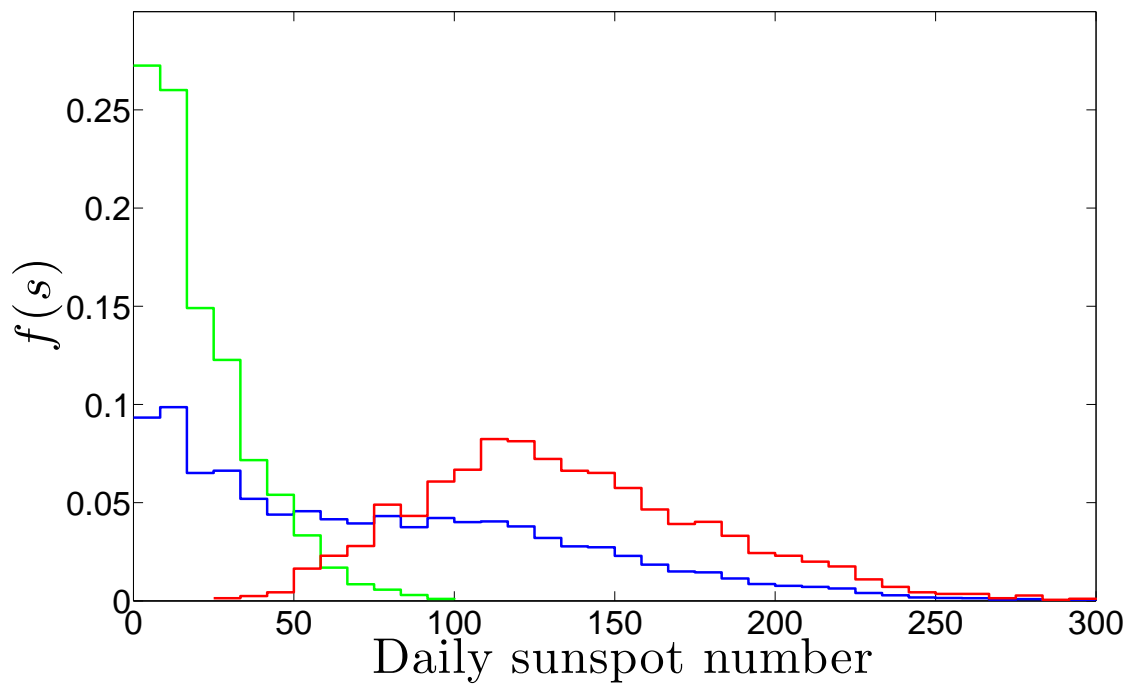}
\caption{Figure showing the time--averaged daily sunspot number distribution $f(s)$ at times of solar minimum (green), solar maximum (red), and the overall time--averaged sunspot distribution (blue) for the last three complete solar cycles (1975--2006).  The distribution is concentrated around zero during solar minimum, and is approximately a positively skewed Gaussian distribution during solar minimum.  The overall distribution for the three cycles is dominated by the large number of days with zero sunspot number. }
\label{fig:DailyTimeAve}
\end{figure}

The model should account for the observed statistical variation in sunspot numbers over a cycle.  The sunspot number distribution $f(s,t)$ changes significantly during a cycle.  Figure \ref{fig:DailyTimeAve} shows the distribution $f(s)$ of daily sunspot numbers averaged in time over sunspot minimums (green), sunspot maximums (red), and the total time--averaged sunspot distribution using daily data for the last three complete solar cycles (1975--2006).  This figure shows that the character of day--to--day fluctuations of the sunspot number is starkly different during different phases of the solar cycle.  The sunspot number distribution during solar minimum (shown in green) is concentrated at zero, and the tail exhibits approximate exponential decay.  The distribution during solar maximum (shown in red) is approximately a positively--skewed Gaussian.  The overall time--averaged distribution during this period (shown in blue) is dominated by the large number of zero sunspot numbers.  The time--averaged distribution is affected by the sampling frequency and number of cycles included.    Section \ref{Section32} shows that the distribution of \emph{monthly} sunspot numbers during 1975--2006 is more strongly bi--modal than the daily distribution shown in Figure \ref{fig:DailyTimeAve}.  The time--averaged distribution of \emph{daily} sunspot numbers for 1850--2010 is approximately exponential.  This is due to the large number of zero sunspot numbers (more than 14\% of days have a zero sunspot number), the large variations in cycle amplitude, and the large fluctuations in maximum sunspot number.  It is this important daily stochastic variation that we are attempting to model.  
  
\subsection{Derivation of a Fokker-Planck equation}
 \label{Section22}
In this section we derive a particular Fokker--Planck equation appropriate to model the sunspot number distribution $f(s,t)$.  The total probability
\begin{equation}
\int_{0}^{\infty} f(s,t) ds
\end{equation}
must always be unity, since probability is a conserved quantity.  Local conservation of 
probability implies the conservation equation 
\begin{equation}
\label{eq:Conservation}
\pd{f(s,t)}{t} = - \pd{F(s,t)}{s},
\end{equation}
where $F(s,t)$ is the probability flux.  A general form for $F(s,t)$ is 
\begin{equation}
\label{eq:GeneralFlux}
F(s,t) = \mu (s,t) f(s,t)- \frac{1}{2}\pd{}{s} \left[\sigma^2 (s,t) f(s,t)  \right]
\end{equation}
where $\mu(s,t)$ and $\sigma^2(s,t)$ are advection and diffusion terms respectively 
\citep{Risken1989}.  The advection coefficient represents the deterministic behaviour of 
the sunspot number evolution (i.e. the effect of the underlying solar cycle on sunspot number), and the diffusion coefficient represents the short--term stochastic behaviour.  

We assume that there is a delayed response to the driver function $\theta(t)$, given by 
an advection term $\mu(s,t)$ in equation (\ref{eq:GeneralFlux}) of the form
\begin{equation}
\label{eq:Advection}
\mu(s,t) = \kappa \left[ \theta(t) - s \right],
\end{equation}
where $1/ \kappa$ is a lag time between the process of driving and the formation of 
sunspots.  When $s<\theta(t)$ the advection term is positive and we expect an increase 
in the sunspot number, and vice versa.  This choice ensures that the sunspot number 
remains close to a level determined by $\theta(t)$.  When the lag time is small the 
sunspot number reacts quickly to changes in the driver.  The driver $\theta(t)$ may be interpreted as a typical sunspot number determined by an underlying model for the solar cycle.

As discussed in section \ref{Section21}, a minimal model of sunspot number variance requires parameters to describe variance at zero sunspot number, and the increase in variance with the increase in sunspot number.  Hence we assume that the diffusion depends quadratically on the sunspot number $s(t)$:
\begin{equation}
\label{eq:Diffusion}
\sigma^2 (s,t) = \beta_0 + \beta_1 s + \beta_2 s^2,
\end{equation}
where $\beta_0,\beta_1$ and $\beta_2$ are positive constants.  
 
With the choices of equations (\ref{eq:Advection}) and (\ref{eq:Diffusion}) the Fokker--Planck equation for the sunspot number distribution is 
\begin{equation}
\label{eq:GeneralFPE}
\pd{f(s,t)}{t} = \frac{1}{2}\pdd{}{s}\left\{ \left[ \beta_0 + \beta_1 s + \beta_2 s^2 
\right]f(s,t) \right\} - \pd{}{s} \left\{ \kappa \left[ \theta(t) - s \right] f(s,t) 
\right\}
\end{equation} 
where $\theta(t)$ is a prescribed driver function.  The initial condition for the PDE 
(\ref{eq:GeneralFPE}) is the delta function
\begin{equation}
\label{eq:IC}
f \left( s,t_0 \right) = \delta \left(s - s_0 \right), 
\end{equation} 
which ensures that total probability is conserved at $t_0$.  As $s \rightarrow \infty$ 
we have the `far field' condition
\begin{equation}
\label{eq:RightBoundary}
f(s,t) \rightarrow 0
\end{equation}
which ensures that very large sunspot numbers are unlikely.  The model has only four 
parameters: the mean reversion $\kappa$; and the three variance terms $\beta_0, \beta_1$ 
and $\beta_2$.  Parameters in the driver function $\theta(t)$ are external to the model.  
As discussed in section \ref{Section21}, we consider this to be the minimum number of parameters required for an accurate description of sunspot data.  The mean reversion represents a time lag in the rise and fall of sunspot numbers associated with changes in the underlying solar cycle.  The three variance parameters represent variance when the sunspot number is zero (one parameter), and the increase of the variance with sunspot number (two parameters).

To determine the behaviour of $f(s,t)$ at $s=0$ we note that the diffusion process which 
underlies the Fokker--Planck equation (\ref{eq:GeneralFPE}) may exhibit complicated 
behaviour near the $s=0$ boundary \citep{KarlinEtAl1981}.  To describe the sunspot 
numbers the underlying Brownian motion must remain non--negative, but there is a 
significant probability of observing a zero sunspot number.  For this reason it is difficult to extend the stochastic differential equation formulation of \cite{AllenEtAl2010} to account for both 
solar maximum and minimum without using an ad--hoc treatment of the stochastic process at zero.  In the Fokker--Planck approach the non--negativity constraint on $s(t)$ means that probability in $s>0$ cannot move into the region $s<0$ and the appropriate boundary condition at $s=0$ is the zero probability flux condition
\begin{equation}
\label{eq:ZeroFluxBC}
\left. \mu (s,t) f(s,t)- \frac{1}{2}\pd{}{s} \left[\sigma^2 (s,t) f(s,t)  \right] 
\right|_{s=0} = 0.
\end{equation}
Although the choice of equation (\ref{eq:ZeroFluxBC}) may appear obvious in the context 
of the Fokker--Planck equation, the formal treatment of the $s=0$ boundary presents a 
problem in the stochastic differential equation approach.  In the Fokker--Planck 
equation formulation however, this physical constraint is a natural component of the 
model.  There are no restrictions on the choice of the driver $\theta(t)$, and during estimation there are no restrictions on the choice of parameters in $\mu(s,t)$.

The time evolution of the sunspot number distribution is defined by the model given by 
equation (\ref{eq:GeneralFPE}), the initial condition (\ref{eq:IC}), the boundary 
conditions (\ref{eq:RightBoundary}) and (\ref{eq:ZeroFluxBC}), and a choice for the driver 
$\theta(t)$. For large $s$ the distribution $f(s,t)$ resembles a positively skewed 
Gaussian distribution.  Near zero sunspot number the zero--flux boundary condition causes probability to accumulate around $s=0$, and $f(s,t)$ often resembles an exponential.  The response of the sunspot number distribution to the driver function is determined by the characteristics \citep{Lindenbaum1996} of the Fokker-Planck equation (\ref
{eq:GeneralFPE}), which are given by the ODE 
\begin{equation}
\label{eq:Characteristic1}
\td{s(t)}{t} = \kappa \theta(t) - \beta_1 - \left( 2 \beta_2 + \kappa \right) s \; \; 
\; \textrm{with}  \; \; s(t_0) = s_0.
\end{equation}
The solution to the characteristic ODE (\ref{eq:Characteristic1}) for the initial 
condition $s_0$ is
\begin{equation}
\label{eq:GeneralCharachteristic}
s(t) = \textrm{e}^{-(2\beta_2 + \kappa)(t-t_0)} \left\{s_0 + \int_{t_0}^{t} \left
[ \kappa \theta(t') - \beta_1 \right]\textrm{e}^{(2\beta_2 + \kappa)t'} dt' \right\}.  
\end{equation}

\subsection{A toy model for a starspot cycle}
  \label{Section23}
We briefly investigate a toy model involving a simple choice for $\theta(t)$ to 
illustrate the features of the model.

Many stars exhibit stellar cycles, so a simple model for the driver function for a 
stellar cycle is the harmonic choice
\begin{equation}
\label{eq:SineDriver}
\theta(t) = \alpha_0 + \alpha_1 \sin (2 \pi t/\alpha_2 + \alpha_3),
\end{equation}
where $\alpha_2$ and $\alpha_3$ determine the period and phase of the cycles, and $
\alpha_0$ and $\alpha_1$ determine the maximum and minimum amplitudes of the driving.  
We assume that our toy stellar cycle involves stochastic formation and decay of 
starspots on the stellar surface, analogous to the Sun.  With the choice of equation 
(\ref{eq:SineDriver}) for a driver the solution to the characteristic ODE (\ref
{eq:GeneralCharachteristic}) has the form
\begin{equation}
\label{eq:CharacteristicSimple}
s(t) = s_{\text{tran}}(t) + s_{\text{per}}(t)
\end{equation}
where
\begin{equation}
s_{\text{tran}}(t) = \left\{s_0 + \frac{\alpha_1 \alpha_2 \kappa}{D}\left[2 \pi \cos 
\alpha_3 - \alpha_2 \left( 2 \beta_2 + \kappa \right) \sin \alpha_3 \right] + \frac
{\beta_1 - \alpha_0 \kappa}{2 \beta_2 + \kappa}  \right\} {\rm e}^{-\left(2 \beta_2 + 
\kappa  \right) t} 
\end{equation}
and
\begin{equation}
s_{\text{per}}(t)  = A_0 + A_1 \sin\left( 2 \pi t/\alpha_2 + A_3 \right),
\end{equation}
with 
\begin{align}
\label{eq:Terms}
D = & \alpha_{2}^{2} \left( 2 \beta_2 + \kappa \right)^{2} + 4 \pi^2 \\
A_0 = & \frac{\alpha_0 \kappa - \beta_1}{2 \beta_2 + \kappa} \\
A_1 = & \frac{\alpha_{1} \alpha_{2} \kappa  }{\sqrt{D}} \\
A_3 = & \tan^{-1} \left[ \frac{\alpha_2 \left( 2 \beta_2 + \kappa \right) \sin \alpha_3 
- 2 \pi \cos \alpha_3}{\alpha_2 \left( 2 \beta_2 + \kappa \right) \cos \alpha_3 + 2 \pi 
\sin \alpha_3} \right].
\end{align}
The term $s_{\text{tran}}(t)$ describes the transient response of the system to the 
initial condition $s_0$, and $s_{\text{per}}(t)$ describes the long--term response to the underlying stellar cycle, which is represented by the sinusoidal driver function.  Specifically, $s(t) \rightarrow s_{\text{per}}(t)$ as $t \rightarrow \infty$.

In equations (\ref{eq:CharacteristicSimple})--(\ref{eq:Terms}), if $\kappa > 0$ and $
\beta_2 \geq 0$ the amplitude of the response is less than the amplitude of the driver, 
with equality achieved in the limiting case where the response time $1/\kappa$ 
approaches zero.  These requirements ensure that the distribution of the sunspot number 
returns to a long--term periodic response to the driver $\theta(t)$ regardless of the 
initial condition $s_0$.  In general there is a lag between the driver $\theta(t)$ and 
the response of the sunspot number, so that the driver and the response are out of phase 
by 
\begin{equation}
\Delta = \alpha_3 - A_3.
\end{equation}
When $1/\kappa \rightarrow 0$ the response to the driver is instantaneous and the phase 
of the driver and reaction coincide, in which case $s(t) = \theta(t)$.  To investigate a 
specific toy model numerically we re--scale time by the period of the star's dynamo, and 
assume our equations are non--dimensional.  The driver function representing the 
periodic variation in the stellar cycles is
\begin{equation}
\label{eq:ToyTheta}
\theta(t) = 1 + \sin \left(2 \pi t \right),
\end{equation}
and the non--dimensional diffusion term representing the stochastic emergence and 
formation of starspots is assumed to be
\begin{equation}
\label{eq:ToyDiffusion}
\sigma^2 (s,t) = 2 + 0.5 s + 0.2 s^2.
\end{equation}
The non--dimensional response time of starspots to the driver is assumed to be 
\begin{equation}
\label{eq:ToyKappa}
1/\kappa = 0.25 \, .
\end{equation}
The Fokker-Planck equation governing the time evolution of the non--dimensional starspot 
number on this star is 
\begin{equation}
\label{eq:ToyFPE}
\pd{f(s,t)}{t} = \frac{1}{2}\pdd{}{s}\left[ \sigma^2(s,t)f(s,t) \right] - \pd{}{s} \left
[ \kappa \left( \theta(t) - s \right) f(s,t) \right]
\end{equation} 
where $\theta(t)$, $\sigma^2(s,t)$, and $\kappa$ are given by equations (\ref
{eq:ToyTheta}), (\ref{eq:ToyDiffusion}), and (\ref{eq:ToyKappa}) respectively.  The 
initial condition is the delta function
\begin{equation}
f(s,t_0) = \delta (s - s_0 ),
\end{equation}
where for simplicity we take $t_0 = 0$.  Equation (\ref{eq:ToyFPE}) describes a dynamic 
system in which the starspot number responds to a simple sinusoidal stellar cycle.  The 
resulting emergence and formation of starspots is stochastic, and the uncertainty 
increases with the starspot number.  The response of the sunspot number distribution to 
the driver $\theta(t)$ is determined by the characteristic curves, which are given by 
equation (\ref{eq:CharacteristicSimple}).

\begin{figure}
\epsscale{1.0}
\plotone{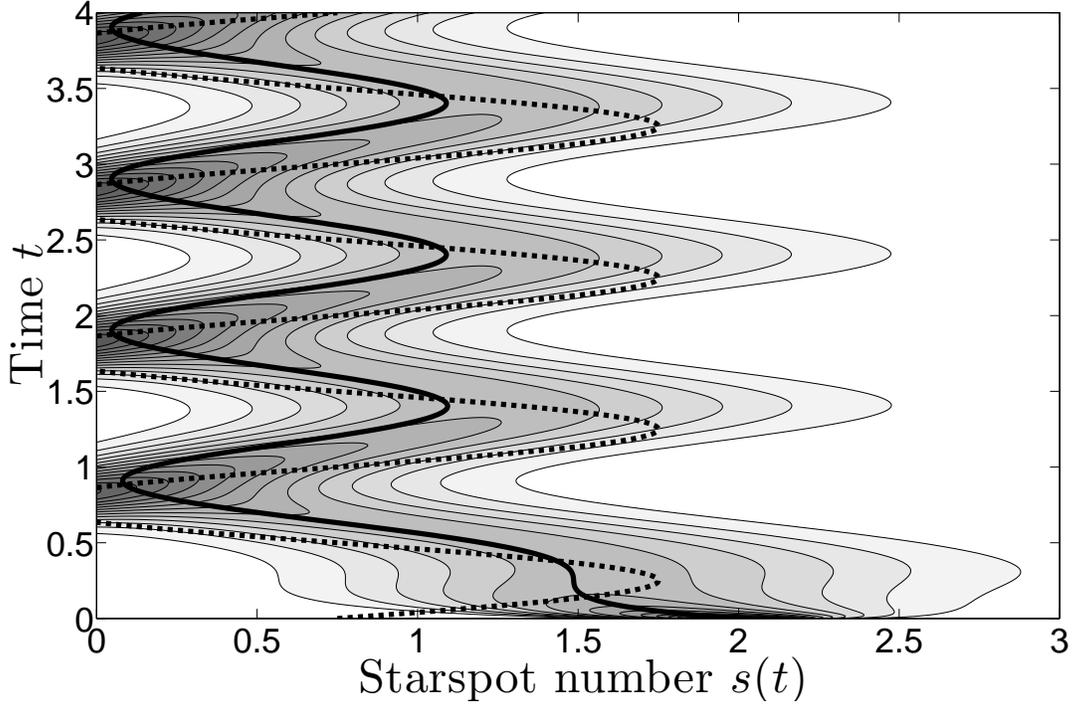} 
\caption{A toy model for a stellar cycle, showing the time evolution of the starspot 
number probability distribution function (pdf) $f(s,t|s_0)$ starting from an initial number $s_0=2$.  The pdf $f(s,t|s_0)$ is shown by the contours.  The distribution follows the characteristic curve (solid line) which is the response to the harmonic driver function (dashed line).  The distribution responds to the initial condition $s_0=2$ before approaching the long--term periodic response given by $s_{\text{per}}(t)$.  The variance of $f(s,t|s_0)$ is greater during stellar maximum.  Probability `accumulates' around zero starspot number during the stellar minimum due to the zero probability flux boundary condition at $s=0$.}
\label{fig:Characteristics2}
\end{figure}

Figure \ref{fig:Characteristics2} shows the numerical solution of the Fokker--Planck 
equation (\ref{eq:ToyFPE}) for the toy model with $s_0=2$.  The driver function (dashed 
curve) and the characteristic (solid curve) are superposed on the contours of the 
distribution in the $s$--$t$ plane.  The distribution exhibits a lag with respect to the 
driver (given by the angle $\Delta = - 0.96$) and follows the characteristic curve from 
the initial condition $s_0=2$ to the long--term response described by $s_{\text{per}}(t)
$ with $A_0 = 0.80$ and amplitude $A_1 = 0.52$.  The figure illustrates the `accumulation 
of probability' about zero starspot number around times of minimum, due to the zero 
probability flux boundary condition at $s=0$.  The variance in the starspot number 
increases with $s(t)$, and the figure shows that the response of the starspot numbers to 
the driving is more varied around starspot maximum.

\section{Parameter estimation}
  \label{Section3}
\subsection{Parameter estimation for diffusion processes}
  \label{Section31}
An important advantage of the Fokker--Planck equation formulation presented here is that 
it allows statistically rigorous estimation of model parameters from data.  The time 
series of sunspot numbers $\tb{s} = \left\{s\left(t_0 \right),s\left(t_1 \right),
\ldots,s\left(t_T \right) \right\}$ is considered to be a discretely observed 
realisation of the underlying continuous diffusion process. The distribution of $s(t)$ 
at time $t_{i+1}$ is dependent only on the previous observation $s_i = s(t_i)$ [the 
Markov property \citep{KaratzasEtAl1991}].  The observations are assumed to be generated 
according to the conditional pdf $f \left(s,t|s_{i};\tb{\Omega}\right)$, which depends 
on a set of parameters  $\tb{\Omega}$ we want to estimate from the observed sunspot 
number time series $\tb{s}$.  The conditional pdf $f \left(s,t|s_{i};\tb{\Omega}\right)$ 
satisfies the Fokker-Planck equation
\begin{equation}
\label{eq:EstimationFPE}
\pd{f\left(s_{},t|s_{i};\tb{\Omega}\right)}{t} = \frac{1}{2}\pdd{}{s}\left[\sigma^2(s,t;
\tb{\Omega})f\left(s_{},t|s_{i};\tb{\Omega}\right) \right] - \pd{}{s}\left[\mu(s,t;\tb
{\Omega}) f\left(s_{},t|s_{i};\tb{\Omega}\right)\right] 
\end{equation}
with initial condition
\begin{equation}
f \left(s,t|s_{i};\tb{\Omega}\right) = \delta\left(s-s_i\right)
\end{equation}
and zero flux condition
\begin{equation}
\left. \mu (s,t;\tb{\Omega}) f(s,t|s_{i};\tb{\Omega})- \frac{1}{2}\pd{}{s} \left
[\sigma^2 (s,t;\tb{\Omega}) f(s,t|s_{i};\tb{\Omega})  \right]\right|_{s=0} = 0.
\end{equation}
The values for the estimated parameters are denoted $\widehat{\tb{\Omega}}$.

Maximum likelihood estimates are considered optimal in the sense that they  are both 
efficient and consistent in large samples \citep{DacunhaCastelleEtAl1986}.  
Qualitatively, this means that as the sample size grows, the probability of a maximum 
likelihood estimator being different to the true parameters converges to zero. Also, as 
the sample size grows the variance of the estimator converges to a theoretical minimum 
value.  The likelihood function $\mathcal{L}$ for a realisation $\tb{s}$ is defined as 
\begin{equation}
\mathcal{L}\left(\tb{\Omega}|\tb{s}\right):=\prod_{i=1}^{i=T}f\left(s_{i}|s_{i-1};\tb
{\Omega}\right),
\end{equation}
where $s_T$ is the final observation in the time series $\tb{s}$, and the maximum 
likelihood estimator $\widehat{\tb{\Omega}}$ is the particular $\tb{\Omega}$ which 
maximises the log-likelihood 
\begin{equation}
\label{LogLikelihood}
\log \mathcal{L}\left(\tb{\Omega}|\tb{s}\right) = \sum_{i=1}^{i=T} \log f\left(s_{i}|s_
{i-1};\tb{\Omega}\right).
\end{equation}
For arbitrary advection and diffusion terms in equation (\ref{eq:EstimationFPE}) and/or 
difficult boundary conditions, general solutions for $f(s,t|s_i)$ are unavailable and 
approximation techniques are required.  Jensen and Poulsen (2002) found that the most 
accurate technique for approximating the unknown distribution involved constructing 
sequences of approximations to $f(s,t|s_0)$ using Hermite polynomial expansions about a 
normal distribution \citep{Ait1999,Ait2002}, followed by direct numerical solution of 
the Fokker-Planck equation \citep{Lo1988}.  \citet{HurnEtAl2007} also found that the two 
most accurate techniques for parameter estimation of diffusion processes involved 
maximum likelihood procedures using these approximations for the unknown pdf $f(s,t|s_i)
$.

In our model, probability accumulates around zero sunspot number at times of minimum due 
to the zero probability flux boundary condition.  As a result an approximation of $f(s,t|
s_0)$ by an expansion about a normal distribution is not always accurate.  Hence we do no use the method of \cite{Ait1999, Ait2002}, but instead apply direct 
numerical solution of the Fokker-Planck equation (\ref{eq:EstimationFPE}) to approximate 
the unknown pdf $f(s,t|s_i)$.  The numerical solutions are obtained using an 
exponentially--fitted finite difference scheme \citep{deAllenEtAl1955,Duffy2006} with 
Rannacher time stepping \citep{Rannacher1984}.  These numerical solutions of equation 
(\ref{eq:EstimationFPE}) are then used to find the maximum likelihood estimates $\widehat
{\tb{\Omega}}$ of the parameters $\tb{\Omega}$ of the sunspot number pdf $f(s,t|s_i)$.  
The optimization of the log--likelihood (\ref{LogLikelihood}) is performed using a 
genetic algorithm based on a routine described by \cite{HauptEtAl2004}.
    
\subsection{Maximum likelihood estimation of the monthly sunspot number}
  \label{Section32}
In this section we apply the model discussed in section \ref{Section:FPE} to the monthly 
sunspot number time series.  To introduce the methodology we use the analysis of the 
toy model in section \ref{Section23} with the sinusoidal driver function (\ref
{eq:SineDriver}) and apply it to the last three cycles of the monthly sunspot numbers 
(1975-2006).  Time is measured in months, and we set $t_0 = 0$ to be January 1975.  
Despite the simple (harmonic) representation of the periodic solar cycle, we 
achieve both qualitative and quantitative agreement between the model distributions and 
the sunspot data.  
\begin{table}[htp]
\caption{Maximum likelihood estimates of the model parameters $\tb{\Omega}$ using 
monthly sunspot data 1975--2006.}
\vspace{0.5cm}
\centering \begin{tabular}{c c c c c c c c}
\tableline\tableline 
$\widehat{\alpha}_0$ & $\widehat{\alpha}_1$ & $\widehat{\alpha}_2$ &
  $\widehat{\alpha}_3$ & $\widehat{\beta}_0$ & $\widehat{\beta}_1$ &
  $\widehat{\beta}_2$  & $\widehat{\kappa}$ \\ [-2ex]
 & & {\scriptsize $\mbox{month}$} & {\scriptsize $\mbox{rad}$} & 
{\scriptsize$\mbox{month}^{-1}$} & {\scriptsize$\mbox{month}^{-1}$} 
& {\scriptsize $\mbox{month}^{-1}$} & 
{\scriptsize $\mbox{month}^{-1}$} \\
\tableline
\normalsize 71.04 & -69.38 & 123.4 & 1.222 & 2547 & 12.65 & 0.53 & 6.51 \\[0.5ex]
 \tableline 
\end{tabular}
\label{table:MLMonthly}
\end{table}

\begin{figure}[htbp]
\epsscale{1.0} 
\plotone{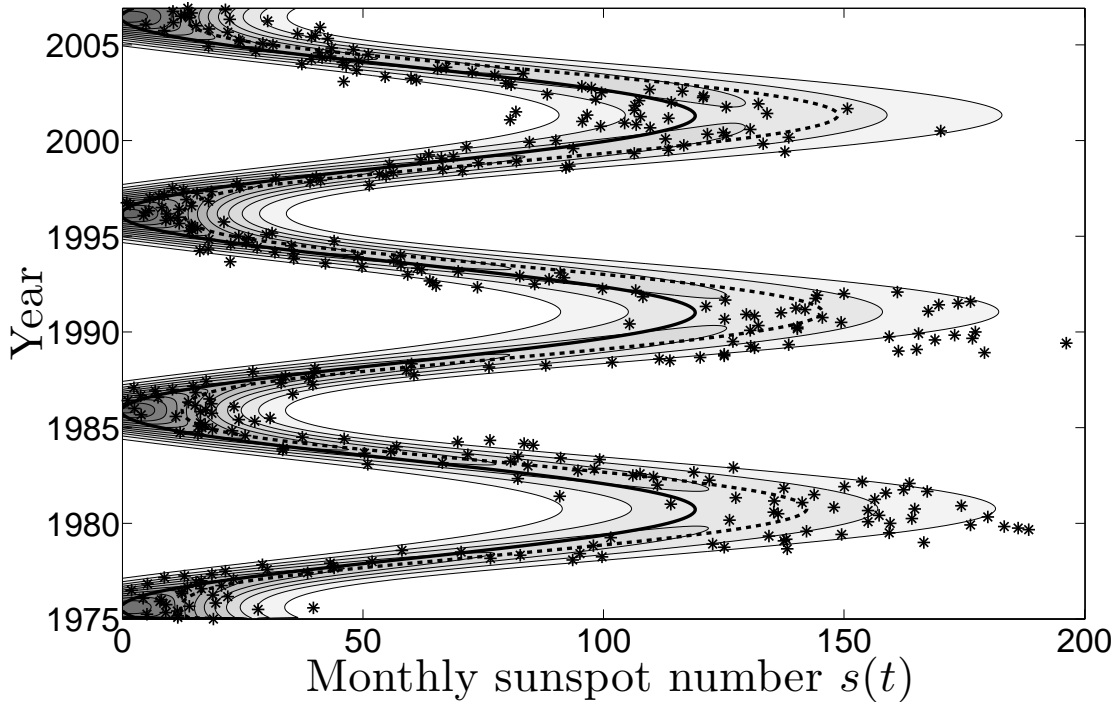}
\caption{Contour plot of the monthly sunspot number distribution found by solving the Fokker--Planck equation (\ref{eq:GeneralFPE}) for the International sunspot data using the 
estimated parameters in Table \ref{table:MLMonthly}.  The initial sunspot number is 
$s_0=18.9$ for January 1975, and the 1975--2006 monthly sunspot data (asterisks) is 
superposed on the contours of the model distribution.  The response to the driver $
\theta(t)$ is given by the characteristic curve (solid), and the expected sunspot number 
$\langle s(t) \rangle$ is the dashed line.  The contours provide a visual representation 
of the shape of the distribution.  Table \ref{table:MonthlyQuantiles1} shows that the 
observed incidence of large sunspot numbers agrees with the tails of the model distribution very accurately.  The amplitude of the response $A_1=-59.67$ is smaller than the amplitude of the driver $\alpha_1 = -69.38$, and there is no noticeable lag.  This plot demonstrates qualitative agreement between the model sunspot number distribution and the monthly data for the years 1975 to 2006.  }
\label{fig:MonthlyEstimation1}
\end{figure}

Table \ref{table:MLMonthly} displays the maximum likelihood estimates of the sunspot 
number parameter set $\tb{\Omega}$ using the monthly data.  Figure \ref
{fig:MonthlyEstimation1} plots the numerical solution of the Fokker--Planck equation 
(\ref{eq:GeneralFPE}) with the maximum likelihood parameter set from Table \ref
{table:MLMonthly}.  The initial observation $s_0 = 18.9$ is for January 1975 and the 
initial condition is the delta function $f(s,0|s_0) = \delta (s-18.9)$.  The final 
observation $s(t_T) = 13.6$ is for December 2006.  The $s_0=18.9$ characteristic curve 
is shown by a solid line, and the monthly data for 1975 to 2006 are superposed on the 
contours of the model sunspot number pdf.  The dashed line is the expected sunspot 
number $\langle s(t) \rangle $, which is defined by
\begin{equation}
\langle s(t)  \rangle = \int_{0}^{\infty} s' f(s',t) ds'.
\end{equation}
The analysis of section \ref{Section23} is appropriate since we are using a harmonic 
driver.  The transient term $s_\text{tran}(t)$ in equation (\ref
{eq:CharacteristicSimple}) vanishes quickly, and the long--term response of the sunspot 
number pdf is determined by the periodic term $s_\text{per}(t)$.  The sunspot number pdf 
fluctuates about the constant $A_0 = 59.42$ and the amplitude of the response $A_1 = 
-59.67$ is smaller than the amplitude of the driver $a_1 = -69.38$.  There is no 
noticeable lag ($\Delta \approx 0$).  Figure \ref{fig:MonthlyEstimation1} demonstrates 
qualitative agreement between the model and the monthly sunspot data, and in particular 
the shape and time variation of the distribution is consistent with the data.  The figure 
illustrates how the characteristic curve determines the long--term response to the 
driver $\theta(t)$, and how the sunspot number varies more during solar maximum.  It 
also shows the accumulation of probability about zero sunspot number at times of solar 
minimum, matching the observed low sunspot number at those times.  

\begin{figure}[ht]
\epsscale{1.0}
\plotone{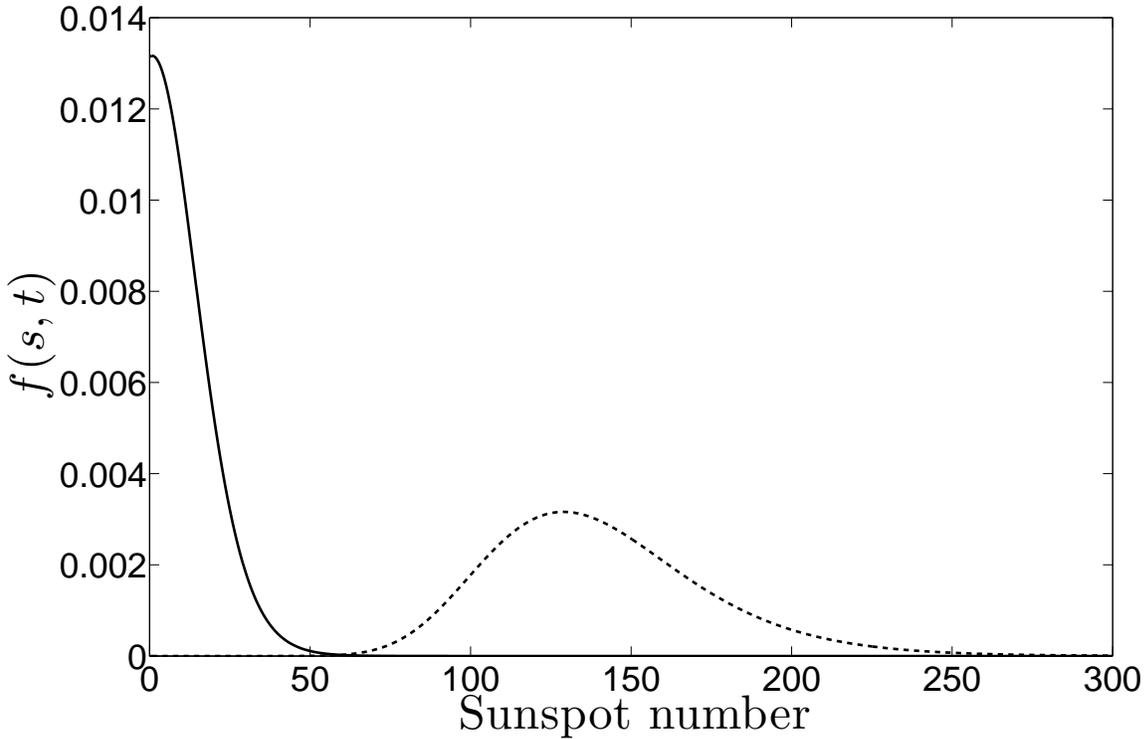}
\caption{Plot of the model sunspot number distribution $f(s,t_\textrm{max})$ at the maximum of cycle 23, and the distribution $f(s,t_\textrm{min})$ at the previous minimum (dashed curve).  The distribution at solar maximum is a positively skewed Gaussian.  The tail of the distribution at solar minimum exhibits exponential decay.  The model distributions qualitatively coincide with the empirical distributions shown in Figure \ref{fig:DailyTimeAve}.}
\label{fig:ModelFst}
\end{figure}

Figure \ref{fig:ModelFst} shows the model sunspot number distribution $f(s,t_\textrm{max})$ at the maximum of cycle 23 (dashed curve), and the distribution $f(s,t_\textrm{min})$ at the previous minimum (solid curve).  The distribution at solar maximum is a positively skewed Gaussian.  The tail of the distribution at solar minimum exhibits exponential decay.  The distributions qualitatively coincide with the empirical distributions in Figure \ref{fig:DailyTimeAve}.

To investigate the statistical agreement between the model and the observations we first 
compare the quantiles of the model and the empirical distribution.  The tails of the 
model distribution represent the probability of observing unusually large or small 
sunspot numbers.  To quantify the accuracy of the tails of the model we calculate the 
lower and upper $a\%$ quantiles $s_L(t)$ and $s_U(t)$ for each month.  These quantiles 
are defined at time $t$ by
\begin{equation}
\label{eq:Quantiles}
 \int_{0}^{s_L(t)} f(s',t|s_0) ds' \; =  \int_{s_U(t)}^{\infty} f(s',t|s_0) ds' = a/100.
\end{equation}
That is, given the initial sunspot number $s_0 = 18.9$ for January 1975, the probability 
of observing a sunspot number less than $s_L(t)$ at time $t$ is $a\%$.  Table \ref
{table:MonthlyQuantiles1} compares the proportion of monthly data lying outside the 
lower and upper $a\%$ quantiles of the model pdf over the period January 1975 to 
December 2006 for $a = 20\%,10\%,5\%,1\%$ and $a=0.50\%$.  Table \ref
{table:MonthlyQuantiles1} shows good agreement between the model values and the 
observations, and confirms that the tails of the model sunspot number distribution are 
accurate over the thirty years.   
\begin{table}[ht]
\caption{Comparison of model and observed tail probabilities for the monthly sunspot 
number for 1975--2006.}
\vspace{0.5cm}
\centering \begin{tabular}{ c c c c c c }
 \tableline\tableline
 Model quantiles & $a=20 \%$ & $a=10 \%$ & $a=5 \%$ & $a=1 \%$ & $a=0.5 \%$   \\  
 \tableline
  Observed upper quantiles & 18  &  9.1  &  4.2  &  0.54  &  0.52 \\ 
  Observer lower quantiles & 23  &  13  &  5.5  &  1.0  &  0.78 \\ 
 \tableline 
\end{tabular}
\label{table:MonthlyQuantiles1}
\end{table}

\begin{figure}[ht] 
\epsscale{1.0}
\plotone{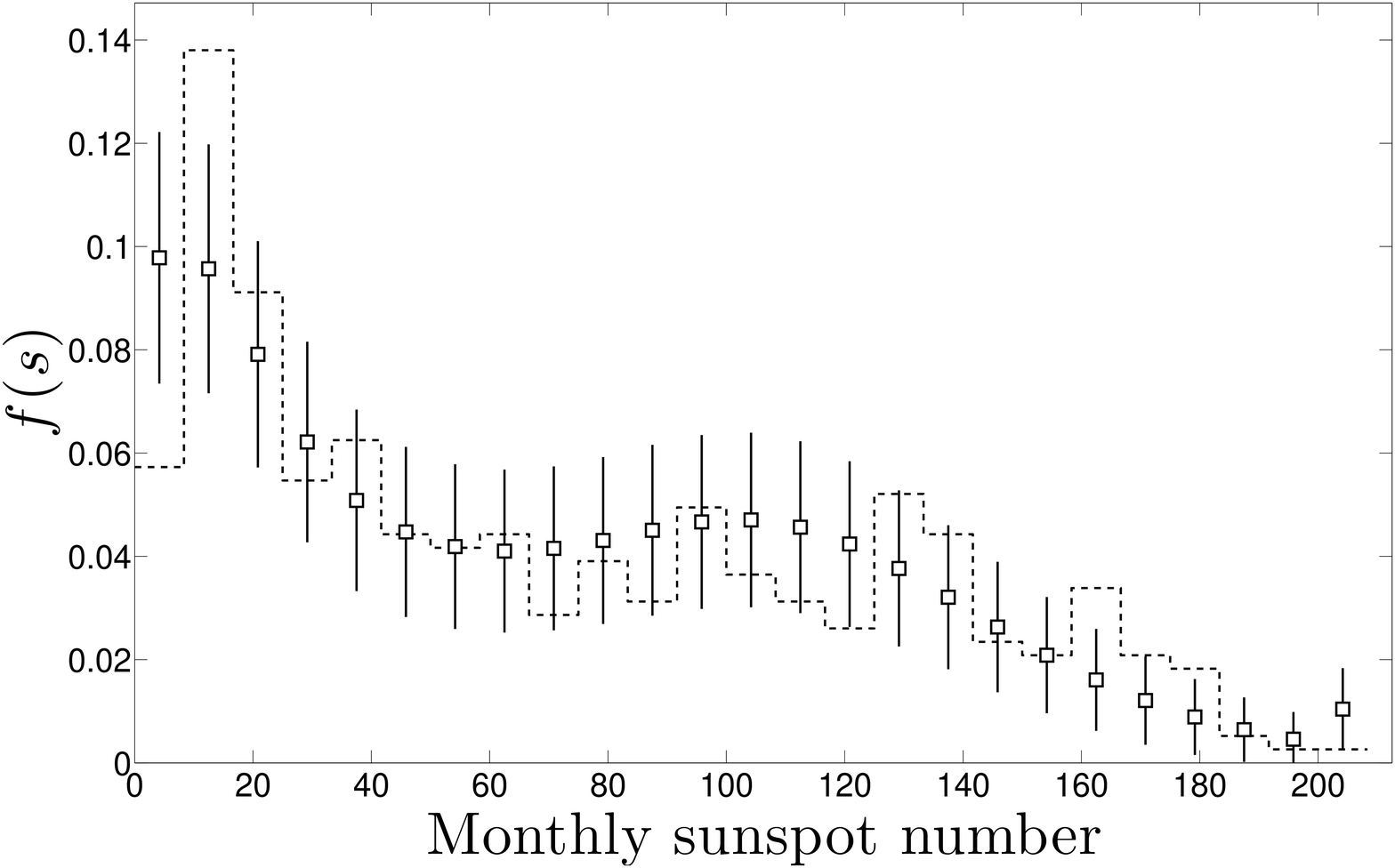} 
\caption{Plot of the model (squares) and empirical (histogram) time--averaged 
distributions for the 1976--2006 monthly sunspot data.  Representative uncertainties are 
shown on the model values.  The time--averaged distribution $f(s)$ is bimodal, with 
modes for both the model and data located at $s \approx 10$ and $s \approx 110$.  A $
\chi^2$ test indicates that the difference between the model and empirical distributions 
is not significant.  This demonstrates quantitative agreement between the model sunspot 
number distribution and the monthly data for the years 1976--2006. }
   \label{fig:TimeAve}
\end{figure}

We also investigate the time--averaged behaviour of the sunspot number distribution over 
a number of cycles, and test the quantitative agreement between the model and data using 
a $\chi^2$ test \citep{PressEtAl1992}.  We construct the time-averaged model distribution
\begin{equation}
\label{eq:EquilibriumDistribution}
f(s) = \frac{1}{t_T - t_0}\int_{t_0}^{t_T}  f(s,t') dt'
\end{equation} 
over the duration of the observations (i.e. $t_0 = $ January 1975 to $t_T = $ December 
2006).  This time--averaged model distribution is calculated by integrating the 
numerical solution to equation (\ref{eq:GeneralFPE}) using the ML parameter set in Table 
\ref{table:MLMonthly} for the interval January 1975 to December 2006.    To calculate 
representative uncertainties from the model distribution we let $O_i$ be the number of 
monthly observations in bin $i$, and $M_i$ be the number implied by the model.  The 
uncertainty in each bin is approximately
\begin{equation}
\sigma_{f,i} \approx \frac{\sqrt{M_i}}{\Delta s \sum_i O_i}
\end{equation}
where $\Delta s = 8.33$ is the bin width.  Figure \ref{fig:TimeAve} compares the time-
averaged distribution (\ref{eq:EquilibriumDistribution}) of the model (squares) with a 
histogram of the monthly sunspot number for the duration.  The representative model 
uncertainties $\sigma_{f,i}$ in each bin are also shown by the error bars.  The model 
distribution reproduces an observed bimodality in the data.  The data shows peaks at $s 
\approx 10$ and $s \approx 110$, which correspond to the minimum and maximum of the 
cycles, respectively.  A $\chi^2$ test \citep{PressEtAl1992} is applied, with
\begin{equation}
\chi^2 = \sum_i \frac{\left(O_i - M_i \right)^2}{O_i + M_i}.
\end{equation}    
The test returns a $p$ value of 0.62, which is not significant.  This says that the data 
cannot be excluded given the model, and indicates quantitative agreement between the 
model and data.

\section{Approximate solution}
\label{Section4}
The maximum likelihood procedure outlined in section (\ref{Section31}) is computationally intensive due to the repeated numerical solution of the Fokker--Planck equation (\ref{eq:GeneralFPE}) inside the log--likelihood (\ref{LogLikelihood}).  In this section we briefly present an analytic approximation which allows parameter estimation of large data sets and/or models where the driver function $\theta(t)$ is difficult to evaluate.  A standard approximation is 
to assume the conditional pdf $f(s,\tau|s_0)$ is approximately normal for small $\tau = 
t-t_0$, so that the log--likelihood can be optimized analytically.  However, this 
approximation is not valid for the sunspot model, since it would imply, for small $s_0$, 
a significant probability of negative sunspot numbers.  Instead we assume that the 
advection and diffusion coefficients $\mu(s,t)$ and $\sigma^2(s,t)$ are constant for 
small $\tau = t-t_0$ and discard the linear terms in the expansions for $\mu$ and $
\sigma^2$, in which case the Fokker--Planck equation is the constant coefficient advection/
diffusion equation
\begin{equation}
\label{eq:AdvDiff}
\pd{f}{t} = \frac{1}{2} \sigma^{2}(s_0,t_0) \pdd{f}{s} - \mu(s_0,t_0) \pd{f}{s}.
\end{equation}   

The solution to equation (\ref{eq:AdvDiff}) with a zero probability flux boundary 
condition at $s = 0$ is
\begin{align}
\label{eq:SunspotSimDist}
f(s,t|s_0) = & \frac{1}{\sqrt{2 \pi \sigma^2(s_0,t_0) \tau}} \left[ \exp \left\{-\frac
{\left[s-\left(s_0+\mu(s_0,t_0)\tau \right)\right]^2}{2 \sigma^2(s_0,t_0) \tau }\right\} 
\right. \nonumber \\
             & \quad \left. +\exp \left \{-\frac{\left[s+\left(s_0+\mu(s_0,t_0)\tau 
\right)\right]^2}{2 \sigma^2(s_0,t_0)\tau} \right\} \right]. 
\end{align}
This solution is analogous to the $O\left(\sqrt{\tau}\right)$ Euler approximation \citep
{KloedenEtAl1999} to the Fokker--Planck equation but with a zero flux boundary condition.  
Equation (\ref{eq:SunspotSimDist}) is the conditional pdf of the random variable $\left|
s(t)\right|$ where $s(t)$ is described by a normal distribution with mean $s_0 + \mu
(s_0,t_0)\tau$ and variance $\sigma^2 (s_0,t_0)\tau$, which we denote
\begin{equation}
\label{eq:SunspotPDF}
s(t)|s_0 \sim \mathbb{N}\left[s_0 + \mu(s_0,t_0)\tau, \sigma^2(s_0,t_0) \tau  \right].
\end{equation}
Equation (\ref{eq:SunspotSimDist}) provides an analytic formula for efficient maximum likelihood estimation involving large data sets, or when $\theta(t)$ is difficult to evaluate.  Equation (\ref{eq:SunspotPDF}) allows simulation of solar cycles for a given driver function $\theta(t)$.  These formulae will be useful in the application of the model to forecasting daily sunspot numbers.   

\section{Discussion}
This paper presents a framework for modelling randomness on top of deterministic models 
of solar cycles in a statistically optimal way.  The Fokker--Planck equation formulation 
allows a general choice of driver function representing the underlying solar cycles, and the framework then describes the stochastic variation in the sunspot number on top of the (assumed) driver.  The approach may be used with a variety of models for variation in solar cycles, including those exhibiting nonlinear and chaotic behaviour.  The model describes a non--negative diffusion process and naturally accounts for the complicated behaviour at the lower boundary at zero sunspot number.  It is therefore valid and useful during both solar 
maximum and minimum.  As such this framework should be particularly useful for solar cycle 
forecasters and is complementary to existing modeling techniques. 

To introduce the methodology, section \ref{Section3} assumes a simple harmonic form for the 
driver function for solar cycles during 1975--2006 (cycles 21--23).  Despite the 
simplification in the description of the periodic variation the model shows both 
qualitative and statistical agreement with the monthly sunspot data.  A $\chi^2$ test confirms consistency between the monthly sunspot data and the model over the three solar cycles.  Further, the model tail probabilities (quantiles) coincide well with the observed rate of occurrence of large and small sunspot numbers.  Since forecasters are largely 
concerned with predicting `abnormally' large events, this is a desirable quality.  The success of the model in reproducing the statistics of observed sunspot numbers despite the use of a simplistic driver function (which has a constant amplitude for three cycles) suggests the importance of short timescale fluctuations to the observed statistics.  

The model neglects an explicit account of the drop in sunspot number associated with regions rotating off the visible disk.  This is a limitation of sunspot data, due to a lack of observations for the reverse side of the sun.  There is no difficult in principle with using data limited in this way: the Fokker--Planck modeling includes this (unphysical) variation in the observed statistics.  In the future a sunspot number for both hemispheres may be available, and the model may be applied to the improved data.
 
The motivation for this model is to provide a statistical description of the large, short--time scale fluctuations in sunspot number, which are important because of the space weather effects produced by large, complex sunspot groups, which may form and evolve rapidly.  This paper has focused on the motivation and formulation of the model, and has demonstrated its ability to reproduce observed sunspot statistics.  In future work we will apply the model in more detail to historical sunspot data and illustrate the utility of the model for forecasting purposes, in particular prediction of cycle 24, the new solar cycle.

% -----------  ACKNOWLEDGMENTS  ------------- %
\acknowledgments
We thank Don Melrose for reading a draft, and an anonymous referee for detailed criticisms which have improved the paper.  Patrick Noble gratefully acknowledges the receipt of a School of Physics Denison Postgraduate Scholarship.

% -----------  FIGURES  ------------- %
% figure 1

% figure 2

% figure 3

% figure 4

% figure 5

% figure 6

% bibliography

\end{document}